# Analogy, mind and life

Vitor Manuel Dinis Pereira
Language, Mind and Cognition Research Group (LanCog).
Philosophy Centre (CFUL).
Faculty of Letters, University of Lisbon
Alameda da Universidade, 1600-214 Lisboa, Portugal
vpereira1@campus.ul.pt

Abstract

I'll show that the kind of analogy between life and information [argue for by authors such as Davies (2000), Walker and Davies (2013), Dyson (1979), Gleick (2011), Kurzweil (2012), Ward (2009)] – that seems to be central to the effect that artificial mind may represents an expected advance in the life evolution in Universe – is like the design argument and that if the design argument is unfounded and invalid, the argument to the effect that artificial mind may represents an expected advance in the life evolution in Universe is also unfounded and invalid.

However, if we are prepared to admit (though we should not do) this method of reasoning as valid, I'll show that the analogy between life and information to the effect that artificial mind may represents an expected advance in the life evolution in Universe seems suggest some type of reductionism of life to information, but biology respectively chemistry or physics are not reductionist, contrary to what seems to be suggested by the analogy between life and information.

Acknowledgements

My mother, Maria Dulce.

# Introduction

The analogy between life and information — for example, pattern recognition, with hierarchical structure and suitable weightings for constituent features (Kurzweil, 2012) — seems to be central to the effect that artificial mind may represents an expected advance in the life evolution in Universe, since information (namely, pattern recognition) is supposed to be the essence of mind and all information (namely, pattern recognition) is implemented by the same basic neural mechanisms. And since we can replicated these mechanisms in a machine, there is nothing to prevent us from set up an artificial mind—we just need to install[1] the right pattern recognizers.

# Artificial mind and cognitive science

The landscape of the artificial mind research can be described as follows: machine learning, reasoning, knowledge representation, restriction fulfilment, search, planning and scheduling, agents, robotics, philosophical foundations, natural language processing, perception and vision, cognitive modelling, knowledge and applications engineering. The main core consists of the first three items: machine learning, reasoning, and knowledge representation. Now let's go to the landscape of the cognitive science research.

---

[1] To create a mind, as argue by Kurzweil (2012), we need to create a machine that recognizes patterns, such as letters and words. Consider: translate a paper. In despite the best efforts to develop artificial universal translators, we are still very far from being able to dispense the human correction of what we write in another language.

Considered the following items. Perception and action, memory, attention and consciousness, the so-called nuclear knowledge, classification, lexicon and ontology, learning, language and representation, choice, rationality and decision, culture and social awareness. The landscape that we can outline with them is the landscape of the cognitive science research, with the artificial mind research as a it's proper part.

Cybernetics, computer sciences, language sciences, neurosciences, brain sciences, psychology, biology, philosophy, mathematics, physics, engineering sciences, in a way all of these sciences contribute to the study of human cognition (the aforementioned items).

The artificial mind research is a way of discovering, describing and modelling some of the main features of consciousness – specifically the cognitive ones. Artificial mind researchers assist cognitive science researchers in explaining how consciousness emerges or could emerge (be caused) by non-conscious entities and processes (a explanatory question), or if consciousness makes any difference for the performance (the operation) of the systems in which consciousness is allegedly present; and if consciousness makes any difference, why and how consciousness makes any difference (a functional question).

A central notion in artificial mind research is that of agent. The idea of an agent is an ongoing and autonomously operating entity in an environment in which there are other processes and agents. We are interested in knowing how an mind agent is designed. Usual questions are the following: how does it perceive, rationalize, decide, learn, how does it perform independently in a mutual environment of problems (specific agents for certain intervention domains)? Artificial mind researchers are interested in multiplying those agents and ask how it works that an enormous variety of those agents can articulate coherently in a multi-agent system (interaction and organization). The combination of these questions (and their answers) can be designated by the term "Distributed Artificial Intelligence" (DAI).

## Consciousness

With respect to consciousness, it can be classified in the following three ways (Block 2002).

1. Access consciousness: we have access consciousness of something if we have its representation, it can be transmitted to each part of the brain and in this way it can be used in our reasoning and (rational) control of our actions. It is likely that this is the type of consciousness that can be implemented in a machine. But we have the problem of debating whether the machine "actually" experiences something or not (and in this case, "actually" is not clearly defined).

2. Phenomenal consciousness: $x$ is in a phenomenal conscious state if $x$ experiences something that characterizes that state. The criterion widely used to talk about phenomenal consciousness is that of "there is something it is like to be in that state". For example, if we are phenomenally conscious of a bright blue sky, then it is because we are experiencing something that makes that mental state a phenomenal conscious state. This experience is the key concept of phenomenal consciousness.

Block identifies the following three differences between access consciousness and phenomenal consciousness.

2.1. Access consciousness is completely defined by a representation (such as a logical agent clause that represents a concept or a fact). Phenomenal consciousness can also have a representational component, but what identifies it is the experience of $x$ (an agent) so that if $x$ were not in this phenomenal conscious state it would not have the experience that it *de facto* has.

2.2. Access consciousness characterizes a mental state as a conscious state because their relations with other modules (in other words, access consciousness uses a functional way of classifying mental states as a conscious states). Being aware is being capable of reasoning and acting, of being stimulated and responding to those stimuli.

2.3. Phenomenal consciousness identifies types of conscious states. For example, all the sensations of pain are phenomenal conscious states of the same type, pain. But if we consider each pain from the perspective of access consciousness, each pain is a different conscious state because it causes different reactions, memories and inferences.

To better illustrate the difference between access and phenomenal consciousness, Block describes cases of the first without the second and vice-versa, of access without phenomenal (a) and phenomenal without access (b) . Those cases are, for example, the following.

(a) An individual can have his visual cortex damaged (have suffered an injury in the V1 area), there are things in his field of vision that he cannot see, the so-called blind spots, and even so respond with elevated exactness to questions concerning the properties of those visual stimuli. This pathology, called blind-sight (Holt, 2003), exemplifies the case of access consciousness without phenomenal consciousness. Phenomenologically, the individual is not aware of anything, but his visual cortex damaged does not preclude this individual from representing those stimuli. His V1 area has been injury, but his representations enable the individual to respond to such visual stimuli. Are their representations that enable him to respond to such visual stimuli.

However, we can still give another example. The one from the Belief-Desire-Intention (BDI) agent (Bratman, 1987), which does not have experiences: he is presumably "aware" of everything in front of him but does not experience any of it [a discussion related to this example is the thought experiment of the Chinese Room by Searle (1980)]. His alleged "consciousness" is presumably "access", not phenomenal consciousness.

(b) One case of phenomenal consciousness without access consciousness is, for example, in which we experience the environmental city sound, because of being so used to living with it, we do not represent it. Perhaps a friend of yours, used to the silence of the countryside, could find it strange how we are able to live in the environmental city sound. The reason is that we are not access conscious even though we are phenomenal conscious of it.

3. Self-awareness: is the state of something when there is an internal representation of oneself. For example, a chimpanzee or a baby (baby around two years, two and one-half years old) is capable of recognizing itself in the mirror but a dog is not. It is likely when a dog looks at the reflected image (of itself) in the mirror, it is conscious of the phenomenal but it interprets the representation to which it has conscious access as another dog.

Coelho (2008) asserts the need of a theory of subjectivity and a theory of the body. The difficulty of the subjectivity theory can be illustrated in the following way: we are not capable of having the sensations of a bat (Nagel, 1974) because we are not bats. And the difficult thing about the theory of the body, in the following way: robotic "organs" are not organs from natural selection, but our brain is an organ of natural selection (Edelman, 2006).

The main difficulty is the so-called phenomenal consciousness. The so-called hard problem of consciousness (Chalmers, 1995). There is nothing we know more intimately than the conscious experience, but there is nothing more difficult to explain. However, this difficult is far from being exclusive to the artificial mind research. For example, in neuroscience the farthest one gets are the neural correlates of access consciousness.

In other words, "access consciousness" refers to the possibility of a mental state to be available to the rest of the cognitive system (to be available, for example, to our production system language like when we try to describe the stinging sharpness of a pinprick, the taste of chocolate or the vibrant red of a fire truck). The access is representational in a way that phenomenology is not: the contrast is between feeling that sting, savoring that chocolate or seeing that red and associated representations such that we may not access these representations (not being in possession of relevant concepts) but, if we experience, we have the experience that in fact we have (for example, see the red of the truck in contrast with see that this truck is red).

In artificial mind research the alleged "consciousness" one gets are also presumably "access", agents have "representations of" their "own internal states", the so-called "self-awareness". Examples of these agents are Homer – implemented by Vere and Bickmore (1990) – and "Conscious Machine" (COMA) – project by Schubert et al. (1993).

Architectures such as State, Operator And Result (SOAR), Intelligent Distribution Agent (IDA) and Adaptive Control of Thought—Rational (ACT-R) are computational models of human cognition (for example, real processing time). However, researchers working in those areas do not explicitly attempt to build an agent with "access consciousness".

Other research projects involve the construction of androids. These have provided an empirical device for various debates: the debate about the relationship between the mind and the body (unifying the psychological and biological), the relationship of the social interaction with internal mechanisms (unifying social sciences and cognitive psychology), the alleged reductionism in neurosciences (the so-called "theories of creation of artificial intelligence"), connectionism versus modularity in cognitive science (the architectures which produce responses similar to human ones), nature versus creation (the relative importance of innateness and learning in social interaction). The construction of androids could very well provide empirical data to the study of subjectivity.

Here, we must note the following: missing a theory of subjectivity is not missing information about subjectivity; rather, this information about subjectivity may be available (presumably provided by researchers in artificial mind) but still lack a theory of subjectivity.

For example, consider what happens when you look at a Necker (1832) cube: suddenly it flips, and although the retinal image and visible two-dimensional (2-D) structure are unchanged, the three-dimensional (3-D) interpretation is different. Lines, or rather cube-edges, that once sloped down away from the viewer now slope up away from the viewer, and the vertical square cube face that was previously farther away is now nearer.

The Necker flip in what it's like to see the pattern of lines as a cube is likely to occur in visually sophisticated robots, under appropriate conditions. There can be no reason for which that variation in what it is like to see these lines as a cube could not take place in robots visually sophisticated (in appropriate conditions).

However, be or not be well informed about this x it is a epistemological problem, not a ontological problem and in this sense, of information not being a ontological problem, Sloman (1996) can say that here — subjectivity — there is none philosophical problem.

We have information about, but not a theory of, subjectivity because here, we are confusing two things: epistemology and ontology. One thing is how we know; another what things are.

Artificial mind research contributes to the study of human cognition and also to the study of subjectivity: contributes not exhausted the study of subjectivity.

The so-called Turing (1950) Test assumed, in its evaluation of intelligence, that the mental does not have to be embodied. However, Turing was wrong regarding the nature of the mental. The so-called Total Turing Test (TTT) preserves the idea that the mental has to be embodied (Harnad, 1991). The candidate to the TTT has to be capable of doing, in the world of objects and persons, all they can do, and do them in a way that is indistinguishable (to people) from their workings. The environment and the set design (Coelho 2008).

So, arguably, we have experimental grounds to build androids. By studying neuron-cognitive mechanisms with the additional and subsidiary help of androids, evaluating their interactions with human beings, researchers can hope to build a bridge, for example, between Neuroscience and the Behavioural Sciences.

With androids we have an experimental apparatus for tests of subjectivity, the phenomenological properties of human bodies being allegedly the same as the phenomenological properties of androids' bodies. Even if androids have not subjectivity, we put our subjectivity — confused by androids' bodies — in androids experimentally, at least by the two seconds (Ishiguro, 2005) of our confusion between androids' bodies and human bodies because of the phenomenological properties of the bodies of the androids.

We need both working hypothesis about the study of the human mind, a theory of subjectivity and a theory of the body. Human beings have the mind that has because they have the body that have, there are no disembodied mind, outside the environment (as instantiated by humans). The mind, even if the mind is a distinct substance from the body, gets most of its stimulation from the body. Furthermore, the mind acts through the body. Given that so much of mental activity arises from bodily stimulation and so much of it is designed to contribute to bodily movement, the human mind is radically unlike, say, the mind of a pure intellect as God (if exist). Taking this seriously, it seems that the human mind could not exist without a body.

The consciousness of human beings is both of access and phenomenal.

Our problem is that there is no place for a necessary connection with physiology in the space of possible development defined by the concept of the mind. Although such conceptual expansion does not imply a contradiction with the essential nature of the subjective experience, nothing precludes a expanded concept of mind from preserving the features of the former concept and allowing the discovery of this connection (Nagel 1998, 2002).

Homer and COMA (only to return to the examples given above), as a working hypothesis about the study of the human cognition, presumably have "access consciousness" ("representations of") but not phenomenal consciousness (subjectivity).

Presumably "access consciousness" of agents as Homer and COMA cannot be separated from a body (Total Turing Test).

However, this body cannot be any aggregate of matter; rather, a body must be indistinguishable from humans to humans: human beings looking at these bodies and confuse us, to "process" (Ishiguro, 2005) them as to other humans.

The phenomenological properties of the bodies of these agents — that is, the way they appear to us — are indistinguishable from the phenomenological properties of human bodies. Our brain "processes" (Ishiguro, 2005) androids (note that the sophisticated robots Sloman 1996 talks about have a body very different from ours) as human for two seconds. There are studies, as Ishiguro 2005, showing that in 70% of participants this is the case. It is for this that we need a theory of subjectivity and a theory of the body as working hypothesis about the study of the human mind.

Notwithstanding, the kind of analogy between life and information [argue for by authors such as Davies (2000), Walker and Davies (2013), Dyson (1979), Gleick (2011), Kurzweil (2012), Ward (2009)] – that seems to be central to the effect that artificial mind may represents an expected advance in the life evolution in Universe – is like the design argument and that if the design argument is unfounded and invalid, the argument to the effect that artificial mind may represents an expected advance in the life evolution in Universe is also unfounded and invalid.

The classic watchmaker analogy

The design argument presented and criticized, for example, by Hume in his Dialogues concerning natural religion (1779), can be formulated as the classic watchmaker analogy as follows.

1. The clock, for its complexity and the way is ordered, is a machine that has to have an intelligent author and builder, with proportional capacities to his work — a human watchmaker.

2. The world, for its complexity and the way is ordered, it is like a clock.

3. Therefore, the world also has to have a smart author and builder, with proportional capacities to his work — the divine watchmaker (God).

Succinctly, this argument holds that given the allegedly similarities between a clock and the world, just as we can assume that an intelligent entity built a clock in a specific way and for a specific purpose, presumably we can do the same for the world. While in the first case, the most plausible hypothesis for the builder of the clock would be a human watchmaker, in the second case, the most plausible hypothesis for the builder of the world would be a "divine watchmaker" because, presumably, only such a being could be capable of such a work.

This argument is an analogy, but, as we shall see next, it raises several problems.

Consider this: it is obvious that the world is complex, has an order and natural events have a regularity, yet the analogy with the watch is fragile, remote and reductive.

The classic watchmaker analogy is fragile, remote and reductive

Firstly, it is fragile, because while the clock is a perfect machine, the world is a "machine" full of imperfections and irregularities that go beyond their usual order or regularity.

Secondly, it is remote, because any similarities between the watch and the world can only be regarded as very distant similarities, only in some aspects. That is, one cannot say with certainty that the world order is similar to the order of the clock, because while we are sure, by experience, that the clock and their order were created according to an end, we have no certainty (not having had any experience of this) that the world and its order were even created, much less that it occurred in accordance with an end (that would be divine) and not just the natural accident (the latter explanation is, moreover, the scientific explanation).

Thirdly, it is a reductive analogy, because while the clock is a machine with a limited complexity to its small dimensions, the world is a "machine" not comparable to the dimensions of the watch, so its complexity cannot be compared with that of the clock.

Now, an analogy can be established from an example that is similar in a relevant aspect — in the case of the watchmaker analogy, the example would be the clock and the relevant aspect would be the complexity of the clock comparable to the complexity of the world. And we have seen that the watchmaker analogy does not fulfil these conditions, so we conclude that the analogy is neither founded nor valid. Therefore, the argument is unfounded and invalid and should not be considered as a good proof of the existence of God.

The analogy between mental life and information is of the same kind of analogy involved in the argument from design.

From the fact that there are mental operations as thought and intention in some parts of nature, particularly in humans and other animals, it does not follow that this may be the rule of the whole that is the nature (that farther exceeds parts as humans and other animals).

The analogy between life and information takes a part (information) by the whole (life).

The idea that a natural biological function of the brain is processing information has not been established empirically by cognitive neuroscience, is a metaphor. The concepts of "processing" and "information" are concepts of folk psychology that seems scientifically rigorous, but are not scientifically rigorous. Concepts as "pattern recognition" does not exhaust all mental activity: if any mental activity falls under the concept of "pattern recognition", is only part of the activity of the mind.

In what way does thinking co-occur with a stimulus and categorizing it? When I am thinking about Waltham (Massachusetts, USA) while in Lisbon (Portugal), I am not recognizing any presented stimulus as Waltham (Massachusetts, USA) — since I am not perceiving Waltham (Massachusetts, USA) with my senses. There is no perceptual recognition going on at all in thinking about an absent object. So concepts as "pattern recognition", although some part of what there is to say about the nature of thought — such as when I am perceiving Waltham (Massachusetts, USA) with my senses — is far from all there is to say about the nature of thought.

Contrast with Pereira (2015), in which the relevant computation to the effect of the occipital and left temporal correlates of the distinction between access and phenomenology is the computation of the high degree of visibility "4" and "5" assigned by the participants in both experiments to the correctly identified stimuli (and what there are more in the second experiment is more incorrect answers than in the first experiment), because to distinguish electrophysiologically the access from phenomenology we need that access and phenomenology will be cognitively consciousness of something and we need that access will be the same for all participants in the two experiments. In all experiments (Pereira, 2015), not only are targets always shown, they must always being shown.

Reach to the explanation of the whole [nature, as in the discussion of the argument from design by Hume; life, as in the discussion of the analogy between life and information by authors such as Davies (2000), Walker and Davies (2013), Dyson (1979), Gleick (2011), Kurzweil (2012), Ward (2009)] starting with just one part (humans and other animals, as in the discussion of the argument from design by Hume; information, as in the discussion of the analogy between life and information), without more, makes these arguments very weak: to the effect of the existence of God (criticized by Hume); to the effect of the analogy between life and information [argue for by authors such as Davies (2000), Walker and Davies (2013), Dyson (1979), Gleick (2011), Kurzweil (2012), Ward (2009)].

At the same time, as Hume says, if we are prepared to admit (though we should not do) this method of reasoning as valid, why then choose the part of nature that says more about us, and not choose another part of nature?

Or, as I says, why then choose the part of mental life that says more about perceptual cases and not emotion, imagination, reasoning, willing, intending, calculating, silently talking to oneself, feeling pain and pleasure, itches, and moods—the full life of the mind? Certainly, they are nothing like the perceptual cases on which the analogy between life and information rest.

Why then choose specifically some of the access features of consciousness and not the phenomenology features of consciousness (Pereira, 2015)? Namely, why then choose the part of mental life that says more about access features of consciousness and not the phenomenology features of consciousness as, for example, Pereira, 2015? Certainly, phenomenology features of consciousness are nothing like (see Pereira, 2015) the information about phenomenology features of consciousness on which the analogy between life and information rest.

According to science, were natural events that, in a succession of chances (without any special or divine plan), although according to the "laws of nature", led to the creation of the world and its existence as we know it.

Thus, even before being able to dream even with Darwinian theories and how they revolutionized scientific knowledge, Hume, through his character Philo, already had an objection to the argument from design that he could not imagine be one of scientific basis against the most devastating effects of such an argument from design — namely the watchmaker analogy.

Indeed, the hypothesis of Hume of a succession of chances, besides being more logical and plausible than the theistic hypothesis, is one that most closely matches Darwinian theories of evolution by natural selection, which would arise a century later (in the 19th century), as well as approaches all subsequent scientific discoveries, not only of biology, but also of chemistry, and physics, regarding the possible certainties we can have about the creation of the Universe.

The analogy between life and information seems suggest some type of reductionism

The analogy between life and information, if we are prepared to admit (suppose you do not agree that the kind of analogy between life and information is like the design argument) this method of reasoning as valid (though we should not do), seems suggest some type of reductionism of life to information.

However, biology respectively chemistry or physics are not reductionist, contrary to what seems to be suggested by the analogy between life and information.

On the biological level, for example, molecular genetics cannot provide a derivation base for evolutionary biology (Lewontin, 1983; Levins, 1968) or even for classical genetics (Kitcher, 1984). Particularly, Kitcher (1984: 350) writes: "the molecular derivation forfeits something important. [...] The molecular account objectively fails to explain because it cannot bring out that feature of the situation which is highlighted in the [biological] cytological story". Richard Lewontin (quoted in Callebaut, 1993: 261), in its turn, claim: "Any textbook or popular lecture on genetics will say: 'The gene is a self-reproducing unit that determines a particular trait in an organism'. That description of genes as self-reproducing units which determine the organism contains two fundamental biological untruths: The gene is not self-replicating and it does not determine anything. I heard an eminent biologist at an important meeting of evolutionists say that if he had a large enough computer and could put the DNA sequence of an organism into the computer, the computer could 'compute' the organism. Now that simply is not true. Organisms don't even compute themselves from their own DNA. The organism is the consequence of the unique interaction between what it has inherited and the environment in which it is developing [cf. Changeux (1985), Edelman (1988a, 1988b)], which is even more complex because the environment is itself changed in the consequence of the development of the organism". So, as exemplified by these two quotes from people working in the field, biology is not reductionist.

Neither chemistry nor physics is reductionist. On the chemical level, for example, the reduction of chemistry to quantum mechanics [Cartwright (1997), Primas (1983)] is a case of failed or incomplete reduction.

And the presumed reductionism in physics is also not more successful than biology or chemistry, on physical level, for example, it is not always possible to combine models of gravitation and electromagnetic forces in a coherent way: they generate inconsistent or incoherent results when applied, for example, to dense matter. This is the main problem currently driving people searching for a unified field theory.

Conclusion

Things in the world are not representational, intentional mental states about them is that they are representational, but phenomenological, physical and functional characteristics of mental states (certain type of nerve cell activation co-occurring with our looking at the world) also are not representational, are sensations and experiences.

Cognitive mental states represent, but sensations not represent anything: if certain things out there stimulate nerve cells, are not these cells that representing things out there to being of such and such a manner.

Semantics is out there, things out there stimulate nerve cells, but the co-occurring configuration of these nerve cells with that stimulation, if claim to be "representational or informational or coding", is just a misuse and overuse of terms like "representation": neurons, their synapses, neurotransmitters, receptors molecular et al. are cellular organisms more than we can access because there is no information or representation about cellular organisms that explain what in fact we felt and experienced.

The idea that neurons (their chemistry and physics) "encode" or represent "information" is wrong. If neurons encode or represent, is starting to take for granted what is intended to show: there is no difference between saying that certain Blood Oxygenation Level Dependent (BOLD) (for example, Ogawa et al., 1992) signal (fMRI) or electroencephalogram (EEG) signal correlates with certain information and saying that certain BOLD (fMRI) or EEG signal is correlated with certain conscious mental states (phenomenal or access). What's there here is question-begging. A fallacy, because they assume "information", they study "consciousness": but someone already showed that neurons encode or represent? Neurons neither encode nor represent anything or nothing: what it is that the human voice encodes or represents? Certain sound waves.

Expressions such as "neural code" are not neurons, are us talking about them. They are to be things out there, they are being represented by us, but they themselves are not representations. Expressions like "information" and "representation" can be eliminated, that what the relevant discipline says about neurons (and related) remains informative. And if "information" is a certain kind of frequency, the frequency is enough!

We telephoned, the listener understands us. But we do not say that the signal between these devices, represent or encode or is information.

A book about oceans is not an ocean: we can bathe ourselves in parts of the ocean without have any concept of "ocean" or of "part", we can see red things without seeing that are red (not having the concept of "red ").

Having information about living organisms does not make this information living organisms – they can be "automata" (Descartes in the second of his Meditations on the First Philosophy, 1641). By definition an artificial, for example, plant (information about the way plants look like) is not a living organism, is not a plant. In the same vein, artificial mind is not mind and can not represent an expected advance in the life evolution in Universe in a way suggest by the analogy between life and information.

However, as a tool, pattern recognition (with the additional and subsidiary help of androids) can help us to have more information about subcortical brain somatic-visceral activity co-occurring with "emotional control" such as anger, fear, lust (may contribute to new treatments and medications for psychiatric disorders and neurobehavioral disorders, see above Introduction), have more information about humans and other animals perceptual cases and have more information about subjectivity. Information, neither unfounded nor invalid analogies as the watchmaker analogy, or fallacies as taking the part by the whole, without some type of reductionism or question-begging fallacies.

For example, given methodological concerns of animal experiments as the problem of disparate animal species and strains, with a variety of metabolic pathways and drug metabolites, lead to variation in efficacy and toxicity or as the problem of length of follow-up before determination of disease outcome varies and may not correspond to disease latency in humans (Pound et al., 2004) and given the third of the four Rs (reduction, refinement, replacement and responsibility) — namely replacement, the use of non-living systems and computer simulation [Schechtman (2002), Hendriksen (2009) and Arora et al. (2011)] — pattern recognition can substitute animals in research (namely, for example, drug research and vaccines).